\documentstyle[12pt]{article}
\title{Model representation for local energy transfer theory of isotropic turbulence}
\author{R.~V.~R.~Pandya\\Department of Mechanical Engineering \\University of Puerto Rico at
Mayaguez \\Mayaguez, Puerto Rico, PR 00681, USA}
\begin{document}

\maketitle

\begin{abstract}
An almost-Markovian model equation is proposed for Fourier modes
of velocity field of isotropic turbulence whose statistical
properties are identical to those governed by equations of Local
Energy Transfer theory of turbulence [McComb {\it et. al.}, J.
Fluid Mech. {\bf 245}, 279 (1992)] compatible with the Kolmogorov
spectrum.
\end{abstract}

\section{Introduction}

An attempt to solve closure problem of fluid turbulence led
Kraichnan to propose Direct Interaction Approximation (DIA)
(\cite{Kraichnan58}, \cite{Kraichnan59}) as a pioneer renormalized
perturbation theory (RPT) followed by other RPTs which have been
reviewed from time to time (\cite{Leslie73}, \cite{ McComb90},
\cite{McComb95}, \cite{Lvov91}, \cite{Lesieur97}). In this paper,
our main concern is with Local Energy Transfer (LET) theory of
isotropic turbulence which is compatible with Kolmogorov spectrum
(\cite{McComb78}, \cite{McComb90}). Based on the Edwards's theory
(\cite{Edwards64}), the LET was proposed by \cite{McComb74} in an
Eulerian framework and, since then, has evolved into a set of
equations comprising of fluctuation-dissipation relation and
equations governing the evolution of two-time and single-time
velocity correlations of isotropic turbulent flow-field
(\cite{McComb78}, \cite{McFS92}). The LET has been remained under
persistent surveillance, especially of McComb and co-workers, for
its performance and accomplishments in cases of isotropic
turbulence and related passive scalar convection (\cite{McS84},
\cite{McSH89}, \cite{McFS92}, \cite{OMQ01}, \cite{MQ03},
\cite{FDB94}, \cite{FD00}). The LET's compatibility with
Kolmogorov spectrum (\cite{McComb90}) despite its failure to
comply with random Galilean invariance (\cite{Kraichnan65a}), its
encouraging performance and computational simplicity relative to
some other RPTs (\cite{McComb90}, \cite{OMQ01}, \cite{MQ03}) are
certain niceties of the LET. Further, a long awaited model
representation for LET, if exists, would establish the fact that
statistical properties predicted by the LET are those of a
realizable velocity field.

The model representations are known to exist for a few RPTs
formulated in Eulerian framework, such as, DIA
(\cite{Kraichnan70a}), and Edwards's extended theory
(\cite{Kraichnan71}). The DIA equations were associated with a
Langevin model equation and an almost-Markovian equation was
suggested by \cite{Kraichnan71} and interpreted as a model
representation for Edwards's theory (\cite{Edwards64}) when
extended to non-steady turbulence cases. Also, model
representation exists for Kaneda's theory (\cite{Kaneda81}) in
mixed Eulerian-Lagrangian framework put forward by
\cite{Kraichnan65a}. In the present work, we suggest an existence
of an almost-Markovian type model representation for the LET
theory.

\section{LET theory equations}
%\label{}
Consider a homogeneous, isotropic, incompressible fluid turbulence
in a reference frame $S$ which is stationary in the laboratory.
The Eulerian turbulent velocity field $u_i({\bf x},t)$ in physical
space (${\bf x}$) and time ($t$), with respect to $S$, can be
expressed in terms of Fourier modes $u_i({\bf k},t)$ by
\begin{equation}
u_i({\bf x},t)=\int d^3{\bf k}u_i({\bf k},t)\exp(i{\bf k}\cdot
{\bf x}),
\end{equation}
and which are governed by the Navier-Stokes equation written in
wavevector (${\bf k}$) and $t$ domain:
\begin{equation}
(\frac{\partial}{\partial t}+\nu k^2)u_i({\bf k},t)= M_{ijm}({\bf
k})\int d^3{\bf p}u_j({\bf p},t)u_m({\bf k}-{\bf p},t).
\label{eq1}
\end{equation}
Here $\nu$ is kinematic viscosity of fluid, inertial transfer
operator
\begin{equation}
M_{ijm}({\bf k})=(2i)^{-1}[k_jP_{im}({\bf k})+k_mP_{ij}({\bf k})],
\end{equation}
the projector $P_{ij}({\bf k})=\delta_{ij}-k_ik_jk^{-2}$, $k=|{\bf
k}|$, and $\delta_{ij}$ is the Kronecker delta. The subscripts
take the values 1, 2 or 3 alongwith the usual summation convention
over repeated subscript. The $u_i({\bf k},t)$ satisfy continuity
condition, {i.e.} $k_iu_i({\bf k},t)=0$ and must satisfy reality
requirement $u_i({\bf k},t)=u_i*(-{\bf k},t)$ where $*$ denotes
the complex conjugate. As there is no uniform velocity in each
realization of isotropic turbulence, $u_i({\bf 0},t)=0$. The
derivation of statistical properties of the velocity field from
equation (\ref{eq1}) poses well known turbulence closure problem
due to nonlinear term presents on the right hand side. The various
RPTs provide ways to tackle the closure problem and discussion
here is focused to the solution provided by the LET theory.

The closed set of LET theory equations (\cite{McComb78}, \cite{
McFS92}, \cite{MQ03}) consists of generalized fluctuation -
dissipation relation for the propagator $H_{in}({\bf k};t,t')$,
and equations governing evolution of two-time velocity correlation
$Q_{in}({\bf k},{\bf k'};t,t')=\langle u_i({\bf k},t)u_n({\bf
k'},t') \rangle$ and single-time velocity correlation $Q_{in}({\bf
k},{\bf k'};t,t)=\langle u_i({\bf k},t)u_n({\bf k'},t) \rangle$ of
the velocity field $u_i({\bf k},t)$ satisfying equation
(\ref{eq1}). For isotropic turbulence, these statistical
properties may be further written as
\begin{equation}
H_{in}({\bf k};t,t')=P_{in}({\bf k})H(k;t,t'),
\end{equation}
\begin{equation}
Q_{in}({\bf k},{\bf k'};t,t')=P_{in}({\bf k})Q(k;t,t')\delta({\bf
k}+{\bf k'}),
\end{equation}
\begin{equation}
Q_{in}({\bf k},{\bf k'};t,t)=P_{in}({\bf k})Q(k;t,t)\delta({\bf
k}+{\bf k'}),
\end{equation}
where $\delta$ represents Dirac delta function. The LET equations
for $H(k;t,t')$, $Q(k;t,t')$ and $Q(k;t,t)$ for isotropic
turbulence may be written as
\begin{equation}
Q(k;t,t')=H(k;t,t')Q(k;t',t'), \,\, \forall t>t' \label{let1}
\end{equation}
\begin{equation}
\left(\frac{\partial}{\partial t}+\nu
k^2\right)Q(k;t,t')=P(k;t,t'), \label{let2}
\end{equation}
\begin{equation}
\left(\frac{\partial}{\partial t}+2\nu
k^2\right)Q(k;t,t)=2P(k;t,t), \label{let3}
\end{equation}
where the inertial transfer term $P(k;t,t')$ is
\begin{eqnarray}
P(k;t,t')=\int d^3{\bf p}L({\bf k},{\bf
p})\Bigl[\int_0^{t'}dsH(k;t',s)Q(p;t,s)Q(|{\bf k}-{\bf p}|;t,s)
\nonumber \\ -\int_0^tdsH(p;t,s)Q(k;t',s)Q(|{\bf k}-{\bf
p}|;t,s)\Bigr]
\end{eqnarray}
and equation (\ref{let1}) represents generalized
fluctuation-dissipation relation. Also
\begin{equation}
L({\bf k},{\bf
p})=\frac{[\mu(k^2+p^2)-kp(1+2\mu^2)](1-\mu^2)kp}{k^2+p^2-2kp\mu}
\end{equation}
and $\mu$ is the cosine of the angle between the vectors ${\bf k}$
and ${\bf p}$. The set of LET equations is compatible with the
Kolmogorov spectrum (\cite{McComb90}) and its predictions are
encouraging when assessed against the direct numerical simulation
results of Navier-Stokes equation (\ref{eq1}) (\cite{McS84},
\cite{ McSH89}, \cite{McFS92}, \cite{MQ03}). Now we pose a
question: Is it possible to obtain a stochastic model equation
whose statistical properties are identical to those as predicted
by the LET equations (\ref{let1})-(\ref{let3})? The answer is
affirmative and the model equation is proposed in the section to
follow.

\section{An almost-Markovian model equation for LET}
Consider an almost-Markovian type equation for $u_i({\bf k},t)$,
written as
\begin{equation}
\Bigl(\frac{\partial}{\partial t}+\nu k^2\Bigr)u_i({\bf
k},t)+\alpha(k,t)u_i({\bf k},t)=b_i({\bf k},t) \label{eq3}
\end{equation}
with the statistically sharp damping function $\alpha({k},t)$ and
white noise forcing term $b_i({\bf k},t)$ having zero-mean. For a
particular choice of $\alpha({k},t)$ and statistical properties of
$b_i({\bf k},t)$, this equation (\ref{eq3}) can recover LET theory
equations. We now obtain that particular choice.

For isotropic turbulence, the evolution of $Q(k;t,t')$ and
$Q(k;t,t)$ as predicted by the model equation (\ref{eq3}) is
governed by the following equations, written as
\begin{equation}
\left(\frac{\partial}{\partial t}+\nu
k^2\right)Q(k;t,t')=-\alpha(k,t)Q(k;t,t') \label{eq4}
\end{equation}
and
\begin{equation}
\left(\frac{\partial}{\partial t}+2\nu
k^2\right)Q(k;t,t)=-2\alpha(k,t)Q(k;t,t)+ B(k,t). \label{eq5}
\end{equation}
Here $B(k,t)$ is statistical property of the white noise forcing
term $b_i({\bf k},t)$ and satisfies
\begin{equation}
\langle b_i({\bf k},t)b_n({\bf k'},t')\rangle =P_{in}({\bf
k})B(k,t)\delta({\bf k}+{\bf k'})\delta(t-t').\label{f2a}
\end{equation}
Comparing equations (\ref{eq4}) and (\ref{eq5}) to the equations
(\ref{let2}) and (\ref{let3}), respectively, suggests
\begin{equation}
\alpha(k,t)=-\frac{P(k;t,t')}{Q(k;t,t')} \label{f1}
\end{equation}
and
\begin{equation}
B(k,t)=P(k;t,t)-P(k;t,t')\frac{Q(k;t,t)}{Q(k;t,t')}.\label{f2}
\end{equation}
It should be noted that the model equation's response function
$H(k;t,t')$ is governed by
\begin{equation}
\left(\frac{\partial}{\partial t}+\nu
k^2\right)H(k;t,t')=-\alpha(k,t)H(k;t,t') \,\, \forall \, t>t'
\label{eq4b}
\end{equation}
and together with equation (\ref{eq4}) suggests
$Q(k;t,t')=H(k;t,t')Q(k;t',t')$ identical to generalized
fluctuation-dissipation relation (\ref{let1}) used in LET. Thus
the equation (\ref{eq3}), along with the expression for $\alpha$
given by (\ref{f1}) and statistical properties of white noise
forcing term satisfying (\ref{f2a}) and (\ref{f2}), is
almost-Markovian model representation which can be associated with
LET theory.

\section{Concluding remarks}

The purpose of this paper has been to propose an almost-Markovian
model representation for local energy transfer (LET) theory. The
proposed Markovian model equation can be used to generate velocity
field whose statistical properties are identical to those of LET
theory. And in that respect, prediction of LET theory are for the
velocity field that is realizable and governed by the Markovian
model equation. It should be worth mentioning here that an another
choice for $\alpha ({k},t)$ and $b_i({\bf k},t)$ resulted in an
almost-Markovian model representation for extended Edwards's
theory (\cite{Kraichnan71}). In fact, the self-consistent
Edwards's theory provided the foundation for LET theory
(\cite{MQ03}). And not to our surprise, both LET and extended
Edwards's theories can be associated with the Markovian type model
equation where generalized fluctuation-dissipation relation is
central.

\underline{Acknowledgements}

I am grateful to Nellore S. Venkataraman for help and
encouragement. I acknowledge the financial support provided by the
University of Puerto Rico at Mayaguez, Puerto Rico, USA.


\begin{thebibliography}{21}
\expandafter\ifx\csname
natexlab\endcsname\relax\def\natexlab#1{#1}\fi

\bibitem[Edwards(1964)]{Edwards64}
{\sc Edwards, S.~F.} 1964 The statistical dynamics of homogeneous
turbulence.
  {\em J. Fluid Mech.\/} {\bf 18}, 239--273.

\bibitem[Frederiksen \& Davies(2000)]{FD00}
{\sc Frederiksen, J.~S. \& Davies, A.~G.} 2000 Dynamics and
spectra of cumulant
  update closure for two-dimensional turbulence. {\em Geophys. Astrophys. Fluid
  Dyn.\/} {\bf 92}, 197--231.

\bibitem[
%Frederiksen {\em et~al.\/}(1994)
Frederiksen {\it et al.}(1994)]{FDB94} {\sc Frederiksen, J.~S.,
Davies, A.~G. \& Bell, R.~C.} 1994 Closure theories
  with non-{G}aussian restarts for truncated two-{d}imensional turbulence. {\em
  Phys. Fluids\/} {\bf 6}, 3153--3163.

\bibitem[Kaneda(1981)]{Kaneda81}
{\sc Kaneda, Y.} 1981 Renormalized expansions in the theory of
turbulence with
  the use of the Lagrangian position function. {\em J. Fluid Mech.\/} {\bf
  107}, 131--145.

\bibitem[Kraichnan(1958)]{Kraichnan58}
{\sc Kraichnan, R.~H.} 1958 Irreversible statistical mechanics of
  incompressible hydromagnetic turbulence. {\em Phys. Rev.\/} {\bf 109},
  1407--1422.

\bibitem[Kraichnan(1959)]{Kraichnan59}
{\sc Kraichnan, R.~H.} 1959 The structure of isotropic turbulence
at very high
  Reynolds numbers. {\em J. Fluid Mech.\/} {\bf 5}, 497--543.

\bibitem[Kraichnan(1965)]{Kraichnan65a}
{\sc Kraichnan, R.~H.} 1965 Lagrangian-history closure
approximation for
  turbulence. {\em Phys. Fluids\/} {\bf 8}, 575--598.

\bibitem[Kraichnan(1970)]{Kraichnan70a}
{\sc Kraichnan, R.~H.} 1970 Convergents to turbulence functions.
{\em J. Fluid
  Mech.\/} {\bf 41}, 189--217.

\bibitem[Kraichnan(1971)]{Kraichnan71}
{\sc Kraichnan, R.~H.} 1971 An almost-{M}arkovian
Galilean-{i}nvariant
  turbulence model. {\em J. Fluid Mech.\/} {\bf 47}, 513--524.

\bibitem[Lesieur(1997)]{Lesieur97}
{\sc Lesieur, M.} 1997 {\em Turbulence in Fluids\/}, 3rd edn.
Dordrecht:
  Kluwer.

\bibitem[Leslie(1973)]{Leslie73}
{\sc Leslie, D.~C.} 1973 {\em Developments in the Theory of
Turbulence\/}.
  Oxford: Clarendon Press.

\bibitem[L'vov(1991)]{Lvov91}
{\sc L'vov, V.~S.} 1991 Scale invariant-{t}heory of
fully-{D}eveloped
  hydrodynamic turbulence - Hamiltonian approach. {\em Phys. Rep.\/} {\bf 207},
  1--47.

\bibitem[McComb(1974)]{McComb74}
{\sc McComb, W.~D.} 1974 A local energy-{t}ransfer theory of
isotropic
  turbulence. {\em J. Phys. A\/} {\bf 7}, 632--649.

\bibitem[McComb(1978)]{McComb78}
{\sc McComb, W.~D.} 1978 A theory of time-{d}ependent isotropic
turbulence.
  {\em J. Phys. A\/} {\bf 11}, 613--632.

\bibitem[McComb(1990)]{McComb90}
{\sc McComb, W.~D.} 1990 {\em The Physics of Fluid Turbulence\/}.
New York, NY:
  Oxford University Press.

\bibitem[McComb(1995)]{McComb95}
{\sc McComb, W.~D.} 1995 Theory of turbulence. {\em Rep. Prog.
Phys.\/} {\bf
  58}, 1117--1206.

\bibitem[
%McComb {\em et~al.\/}(1992)
McComb {\it et al.}(1992)]{McFS92} {\sc McComb, W.~D., Filipiak,
M.~J. \& Shanmugasundaram, V.} 1992 Rederivation
  and further assesment of the {LET} theory of isotropic turbulence, as applied
  to passive scalar convection. {\em J. Fluid Mech.\/} {\bf 245}, 279--300.

\bibitem[McComb \& Quinn(2003)]{MQ03}
{\sc McComb, W.~D. \& Quinn, A.~P.} 2003 Two-{p}oint, two-{t}ime
closures
  applied to forced isotropic turbulence. {\em Physica A\/} {\bf 317},
  487--508.

\bibitem[McComb {\it et al.}(1984)]{McS84}
{\sc McComb, W.~D. \& Shanmugasundaram, V.} 1984 Numerical
calculations of
  decaying isotropic turbulence using the {LET} theory. {\em J. Fluid Mech.\/}
  {\bf 143}, 95--123.

\bibitem[
%McComb {\em et~al.\/}(1989)
McComb {\it et al.}(1989)]{McSH89} {\sc McComb, W.~D.,
Shanmugasundaram, V. \& Hutchinson, P.} 1989 Velocity
  derivative skewness and two-{t}ime velocity correlations of isotropic
  turbulence as predicted by the {LET} theory. {\em J. Fluid Mech.\/} {\bf
  208}, 91--114.

\bibitem[
%Oberlack {\em et~al.\/}(2001)
Oberlack {\it et al.}(2001)]{OMQ01} {\sc Oberlack, M., McComb,
W.~D. \& Quinn, A.~P.} 2001 Solution of functional
  equations and reduction of dimension in the local energy transfer theory of
  incompressible, three-{d}imensional turbulence. {\em Physical Review E\/}
  {\bf 63}, 026308--1--026308--5.

\end{thebibliography}
\end{document}